\documentclass[12pt,a4paper,final]{iopart}
\usepackage{iopams}
\usepackage{float}
\usepackage{graphicx}
\usepackage[breaklinks=true,colorlinks=true,linkcolor=blue,urlcolor=blue,citecolor=blue]{hyperref}
\usepackage{tabularx}

\begin{document}

\title[]{Investigation of the physical properties of the tetragonal CeMAl$_{4}$Si$_{2}$ (M = Rh, Ir, Pt) compounds}
\author{N. J. Ghimire, F. Ronning, D. J. Williams, B. L. Scott, Yongkang Luo, J. D. Thompson, E. D. Bauer}
\address{Los Alamos National Laboratory, Los Alamos, New Mexico 87545, USA}
\
\ead{nghimire@lanl.gov}

\begin{abstract}

The synthesis, crystal structure, and physical properties studied by means of x-ray diffraction, magnetic, thermal and transport measurements of CeMAl$_{4}$Si$_{2}$ (M = Rh, Ir, Pt) are reported, along with the electronic structure calculations for LaMAl$_{4}$Si$_{2}$ (M = Rh, Ir, Pt). These materials adopt a tetragonal crystal structure (space group \emph{P4/mmm}) comprised of BaAl$_4$ blocks, separated by MAl$_2$ units, stacked along the $c$-axis.  Both CeRhAl$_{4}$Si$_{2}$ and CeIrAl$_{4}$Si$_{2}$ order antiferromagnetically below $T_{N1}$=14 and 16 K,  respectively, and undergo a second antiferromagnetic transitition at lower temperature ($T_{N2}$=9 and 14 K,  respectively).   CePtAl$_{4}$Si$_{2}$ orders ferromagnetically  below  $T_C$ =3 K with an ordered moment of $\mu_{sat}$=0.8 $\mu_{B}$ for a magnetic field applied perpendicular to the $c$-axis.  Electronic structure calculations reveal quasi-2D character of the Fermi surface.

\end{abstract}
%\pacs{00.00, 20.00, 42.10}

\vspace{2pc}
%\noindent{\it Keywords}: Article preparation, IOP journals
% Uncomment for Submitted to journal title message
%\submitto{\JPA}
% Comment out if separate title page not required
%\maketitle
%\ioptwocol

\section[S1]{Introduction}

Cerium-based intermetallic compounds have received widespread attention in recent years due to the wide variety of strongly correlated electron behavior they exhibit including heavy fermion behavior, quantum criticality, unconventional superconductivity, and complex magnetic order \cite{Lohneysen2007,Stockert2011}.  Their ground state properties are determined by a delicate balance of competing Kondo and RKKY interactions \cite{Doniach1977}, which tend to suppress or promote magnetic order, respectively.  Both interactions arise from the coupling of Ce 4f electrons to conduction electrons.  When the hybridization is sufficiently strong to produce comparable Kondo and RKKY interaction strengths, magnetic order is suppressed to T=0 K.  Quantum fluctuations of the 2nd-order magnetic transition at this quantum critical point (QCP) give rise to unusual power law (or logarithmic) temperature dependencies of the electrical resisitivity ($\rho \sim T^n$ with $n< 2$) and thermodynamic  (Sommerfeld coefficient $C/T \sim -lnT$) properties, in contrast to the expected behavior for a Fermi liquid ($\rho \sim T^2$ and $C/T \sim$ const.)  \cite{Lohneysen2007,Stockert2011}. Unconventional superconductivity often appears near this quantum critical point and  may be mediated by the magnetic fluctuations that are most abundant at the QCP \cite{Monthoux2007}.  The concept of a quantum critical point provides an appealing framework with which to understand the behavior of several classes of correlated electron materials including the f-electron heavy fermion compounds, and the high-$T_c$ iron-pnictide and cuprate superconductors, but the nature of the quantum criticality and its connection to unconventional superconductivity is still fiercely debated.

The f-electron/conduction electron hybridization that controls the behavior of Ce-based compounds depends sensitively on the crystal chemical environment.  Thus, certain families of cerium materials that crystallize in a particular structure type or are made from similar atomic building blocks tend to show interesting correlated electron behavior.  For instance, the chemical environment of the tetragonal CeM$_{2}$X$_{2}$ compounds changes the hybridization to produce a variety of ground states ranging from magnetically ordered (e.g., CeCu$_{2}$Ge$_{2}$) \cite{deBoer1987} to quantum critical (e.g., CeNi$_{2}$Ge$_{2}$) \cite{Kuchler2003}, to superconducting (e.g., CeCu$_{2}$Si$_{2}$) \cite{Steglich1979}; furthermore, the hybridization in a given material may be tuned with modest amounts of pressure or chemical substitution to reach the quantum critical point \cite{Mathur1998,Stockert2011}.  Likewise, the  tetragonal Ce$_{m}$$M_{n}$In$_{3m+2n}$ family,  where M is Co, Rh or Ir, \emph{m} and \emph{n} are the number of the CeIn$_{3}$  and MIn$_{2}$ building blocks, respectively, are prototypical heavy fermion superconductors near an antiferromagnetic quantum critical point \cite{Thompson2006,Thompson2003}. In our search for new families of heavy fermion compounds that may show interesting behavior, we focus our attention on  CeMAl$_{4}$Si$_{2}$  (M=Rh, Ir, Pt). These compounds adopt a tetragonal crystal structure and are the \emph{n} = 1 members of the CeM$_{n}$Al$_{2n+2}$Si$_{2}$  family which, in turn, is  a subfamily of an homologous series of compounds R(AuAl$_{n}$)Au$_{n}$(Au$_{x}$Si$_{1-x}$)$_{2}$ \cite{Latturner2008}, where R is a rare earth element and \emph{n} is an integer. The general formula reflects the fact that the Si site can be fully occupied by Si atoms only (\emph{x}=0) or can be partially occupied. A structure of a member of this family is composed of two building blocks - an R containing BaAl$_{4}$-type quasi-2D layer and a transition metal containing AuAl$_{2}$-type block alternately stacked along the \emph{c} axis as depicted in figure \ref{fig1}(a). Increasing \emph{n}, increases the thickness of  the AuAl$_{2}$ type block which increases the distance between the  adjacent R layers along the \emph{c} axis. Members of the  \emph{x} = 0.5 subfamily  with \emph{n}=0 - 5  have been reported \cite{Latturner2008,Latturner2003}.

Herein, the crystal growth, structure, magnetic, thermal and transport properties of the  (\emph{n}=1) members of the Ce based \emph{x} = 0 sub-family-CeMAl$_{4}$Si$_{2}$ with M= Rh, Ir and Pt are reported.  Electronic structure calculations of the iso-structural non-magnetic La-analogs were also carried out.

\section[S2]{Experimental details}

Single crystals of CeMAl$_{4}$Si$_{2}$  (M= Rh, Ir, Pt) were grown out of Al/Si eutectic flux, similar to the method proposed in Ref. \cite{Maurya2014}. CeMSi$_{3}$ was first prepared by arc melting under an argon atmosphere. The arc melted CeMSi$_{3}$ was mixed with Al$_{88}$Si$_{12}$ in a ratio of 1:8 by weight. The starting materials were loaded into a 2 ml crucible which was then sealed in a silica ampoule under vacuum. The ampoule was heated to 1100 $^{\circ}$C in 6 hours, homogenized at 1100 $^{\circ}$C for 24 hours and then slowly cooled to 700 $^{\circ}$C at the rate of 4 $^{\circ}$C per hour. Once the furnace reached 700 $^{\circ}$C, the excess flux was decanted from the crystals using a centrifuge. Plate like crystals as large as 3 $\times$ 2 $\times$ 1 mm$^{3}$ were obtained. Crystals of iso-structural  analogs LaMAl$_{4}$Si$_{2}$  were synthesized by a similar method.

The structure of the CeMAl$_{4}$Si$_{2}$ crystals was determined by single crystal x-ray diffraction at room temperature using a Burker SMART APEX II charged coupled-device (CCD) diffractometer. The instrument was equipped with a graphite monochromatized Mo K-$\alpha$ x-ray source ($\lambda$=0.71073 \AA). The crystals were mounted on glass fibers using epoxy. A hemisphere of data was collected in single crystal x-ray diffraction using $\omega$ scans, with 10 second frame exposure and 0.3$^{\circ}$ frame widths. Data collection and initial indexing and cell refinement were handled using APEX II \cite{APEXII} software. Frame integration, including Lorentz-polarization corrections, and final cell parameter calculations were carried out using SAINT$^{+}$ \cite{SAINT} software. The data were corrected for absorption using the SADABS \cite{SADABS} program. Decay of reflection intensity was monitored via analysis of redundant frames. The refinement was initiated using coordinates from isostructural literature compounds \cite{Maurya2014}. Structure solution, refinement, graphics, and creation of publication materials were performed using SHELXTL \cite{SHELXTL}.  Verification of the determined structure of the RMAl$_{4}$Si$_{2}$ (R=La, Ce; M= Rh, Ir and Pt) crystals was carried out by Rietveld refinement of the x-ray powder pattern collected on powdered single crystals. The Rietveld refinement was carried out using FullProf \cite{FullProf}. The powder x-ray diffraction patterns of CeMAl$_{4}$Si$_{2}$ are shown in Fig. \ref{fig1b}. The chemical composition of the single crystals was studied using a scanning electron microscope (SEM) with an energy dispersive x-ray spectrometer (EDS). The atomic percentages of Ce, M, Al and Si in CeRhAl$_{4}$Si$_{2}$ and CeIrAl$_{4}$Si$_{2}$ were found  to be 12.3, 12.6, 50.2, 24.7 and 12.6, 12.2, 50.6, 24.6, respectively, which is within the expected uncertainty of about 3-5 \% for the standardless measurements of the composition of  Ce:M:Al:Si = 1:1:4:2 normalized to Ce. A similar result was found in the cases of  LaRhAl$_{4}$Si$_{2}$ and LaIrAl$_{4}$Si$_{2}$. The atomic percentages of  Ce, Pt, Al, Si were, however, found to be 12.0, 15.5, 51.5, 20.2 for  CePtAl$_{4}$Si$_{2}$ corresponding to the Ce:Pt:Al:Si ratio of 1:1.29:4.29:1.68 normalized to the Ce composition. It is consistent with the result obtained in x-ray diffraction presented in section \ref{S3} that the 2h site has a mixed occupancy of 87\% Si and 13\% Pt giving the empirical formula CePt$_{1.26}$Al$_{4}$Si$_{1.74}$ . A similar composition was found in case of LaPtAl$_{4}$Si$_{2}$.

DC magnetization measurements were performed using a Quantum Design magnetic property measurement system (MPMS) between 1.8 K and 350 K and in magnetic fields up to 60 kOe.  Specific heat and resistivity measurements were conducted in a Quantum Design physical property measurement system (PPMS).  Electrical contacts for resistivity measurements were made by spot welding 25 $\mu$m diameter Pt wires onto the sample. The magnetic and resistivity measurements were conducted on oriented crystals.

Electronic structure calculations within density functional theory \cite{DFT} were performed using the WIEN2k code \cite{WIEN2K}. We used the Perdew-Burke-Ernzerhof exchange-correlation potential based on the generalized gradient approximation \cite{PBE}, and included spin-orbit interactions through a second variational method.

\section{Results and discussion}\label{S3}

\begin{figure}[ht]
\begin{center}
  \includegraphics[scale=1]{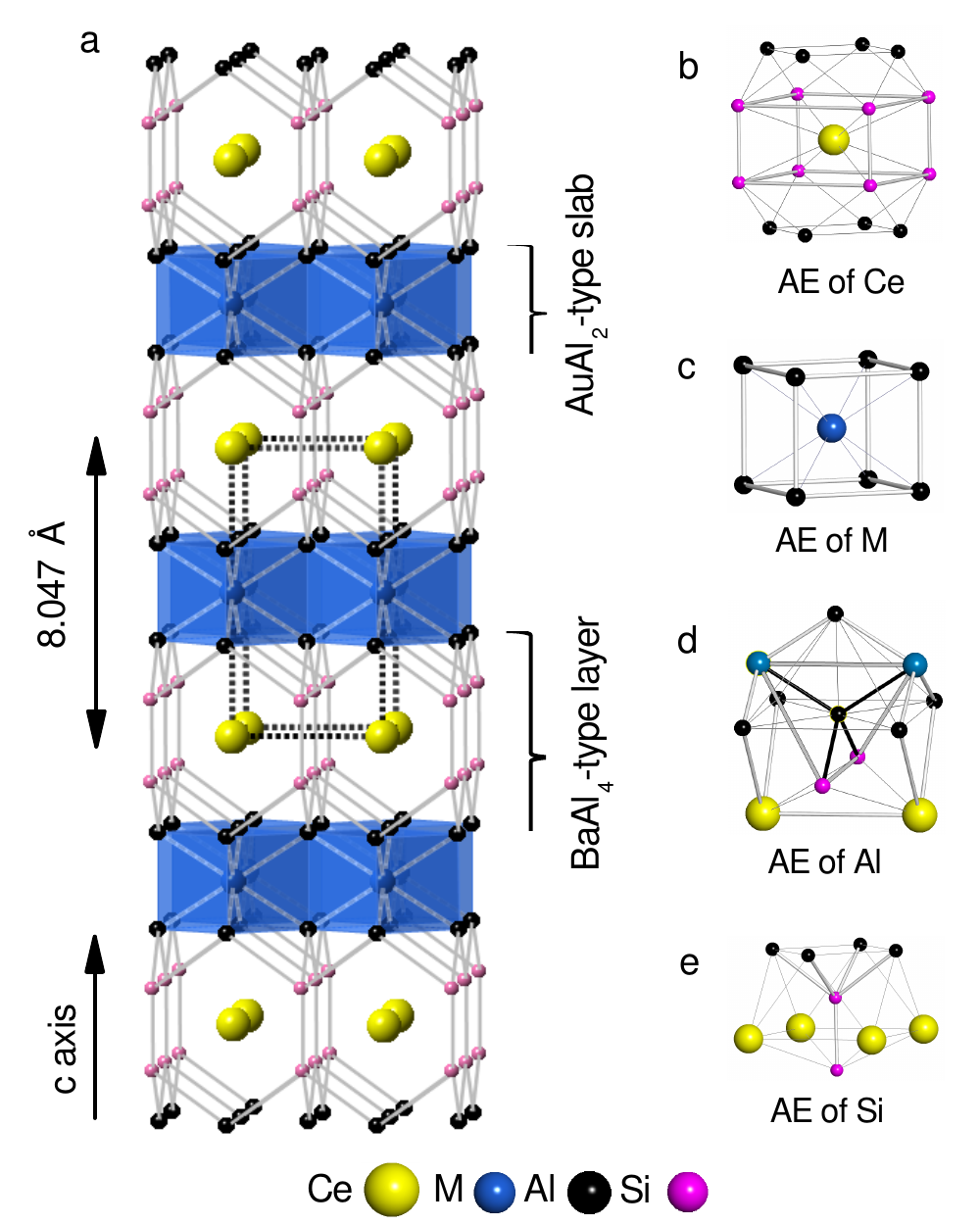}
  \caption{ (Color online) a) Crystal structure of CeMAl$_{4}$Si$_{2}$ showing the stacking of Ce-containing BaAl$_{4}$-type layer and M-containing AuAl$_{2}$-type slab along the \emph{c} axis. Atomic environment (AE) of b) Ce c) M d) Al e) Si.}\label{fig1}
  \end{center}
\end{figure}
\begin{figure}[ht]
\begin{center}
  \includegraphics[scale=.7]{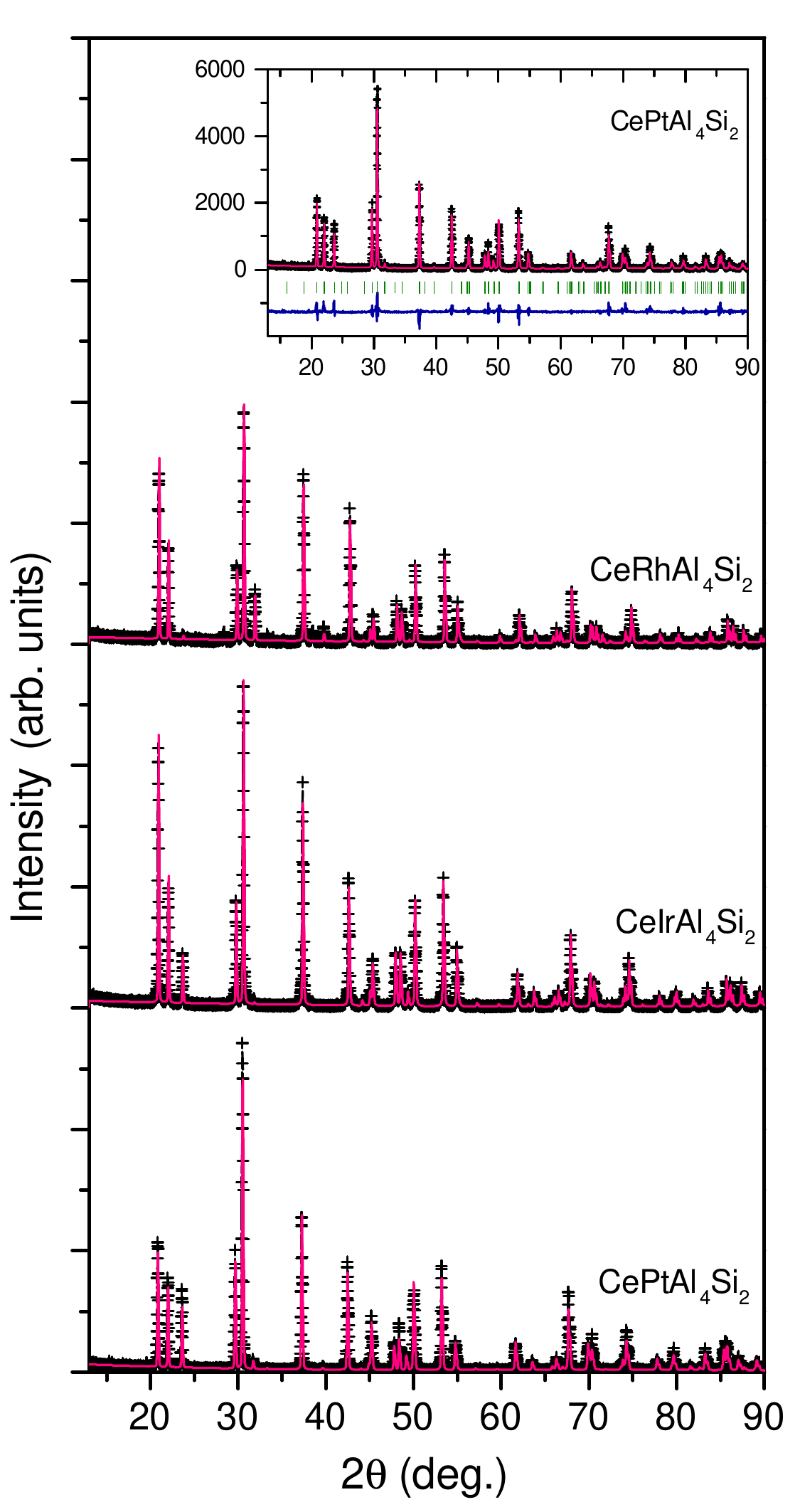}
  \caption{(Color online) X-ray powder patterns of CeMAl$_{4}$Si$_{2}$. The solid line with cross symbols (black) is the data and the solid line (pink) is the calculated powder pattern for each compound. The inset shows the Rietveld refinement of x-ray powder pattern of CePtAl$_{4}$Si$_{2}$. In the inset, the bottom solid line (royal blue) is the difference curve and the middle bars (green) are the calculated peak positions.} \label{fig1b}
  \end{center}
\end{figure}
The CeMAl$_{4}$Si$_{2}$  compounds crystallize in the KCu$_{4}$S$_{3}$ structure type with  space group \emph{P4/mmm}, as shown in Fig. \ref{fig1}(a). The crystallographic data are presented in Table \ref{Table1}. Fractional atomic coordinates and equivalent displacement parameters are given in Table \ref{Table2}. In CePtAl$_{4}$Si$_{2}$, the 2h site (0.5, 0.5, \emph{z}) was refined with joint occupancy of Si and Pt. The 2h site occupancies of Si and Pt were constrained to a total of 1.0 and refined to 0.132(2) for Pt, giving the actual stoichiometry of CePt$_{1.26}$Al$_{4}$Si$_{1.74}$. For simplicity, we use the formula CePtAl$_{4}$Si$_{2}$ through out this article, though the precise formula was used in the analysis of the magnetic and thermal properties measurements. This structure is similar to the disordered $RM_{2}$Al$_{4}$Si, where the Si site has approximately a 50/50 occupancy of Si and M atoms as described in reference \cite{Latturner2008}. In some cases, an ordered variant of RM$_{2}$Al$_{4}$Si has also been reported where Si and M on the Si site are fully ordered making a superstructure with lattice parameter $a^{\star}$ = $\sqrt{2}a$ and a lower symmetry of $P4/nmm $. A fully ordered  KCu$_{4}$S$_{3}$ type structure has been reported for EuMAl$_{4}$Si$_{2}$ with M = Rh and Ir \cite{Maurya2014}. The structure of these compounds may be described as having two basic units that stack sequentially one after another [Fig. \ref{fig1}(a)]. The Ce atoms are surrounded by 8 nearest neighbor Si atoms at 3.2 \AA\ that form a square net in the \emph{ab} plane and 8 Al atoms at a slightly longer distance (3.4 \AA) comprising the atomic environment around Ce with a coordination number of 16 [Fig. \ref{fig1}(b)]. This Ce environment is equivalent to that of  barium in BaAl$_{4}$. This Ce-containing BaAl$_{4}$-type layer alternates with an antifluorite-type transition metal (M) MAl$_{2}$ slab in which the M atoms are coordinated by  a cube of eight Al atoms [Fig. \ref{fig1}(c)]. The local environments of Al and Si atoms are shown in Fig. \ref{fig1}(d-e), respectively. The Ce-Ce distance in the \emph{ab} plane is 4.3 \AA,  and is  8 \AA\ along the c-axis, albiet separated by the MAl$_{2}$ slab.

\begin{table}
\caption{Crystallographic data of CeMAl$_{4}$Si$_{2}$}\label{Table1}
\begin{center}
\begin{tabular}{lccc}
\hline
Compound	       &   CeRhAl$_{4}$Si$_{2}$	  &	   CeIrAl$_{4}$Si$_{2}$	&	CePtAl$_{4}$Si$_{2}$\\
\hline							
Crystal structure & \multicolumn{3}{c}{    	             tetragonal	             }   \\
Space group     	 &		             \multicolumn{3}{c}{    	             P4/mmm	             }            \\
a (\AA)            &	4.227 	          &	  4.233 	            &	4.271 	     \\
c (\AA)	                 &	8.047 	      &	  8.035 	            &	8.060 	     \\
Empirical formula	 &	CeRhAl$_{4}$Si$_{2}$  &	   CeIrAl$_{4}$Si$_{2}$	&	CePt$_{1.26}$Al$_{4}$Si$_{1.74}$	                                                              \\
\hline							
\end{tabular}
  \end{center}
\end{table}

\begin{table}[htdp]
\caption{Fractional Atomic Coordinates and Equivalent Displacement Parameters for CeRhAl$_{4}$Si$_{2}$, CeIrAl$_{4}$Si$_{2}$ and CePtAl$_{4}$Si$_{2}$, Tetragonal, P4/mmm, Z=1}\label{Table2}
\begin{center}
\begin{tabular}{cccccccc}								
		\hline													
Compound	            & atom               &	Wycoff site	&	Occ. &	x	&	y	&	z	      &	U$_{eq}$(\AA$^{2}$)	\\
		\hline													
CeRhAl$_{4}$Si$_{2}$	&	Ce	             &	1b	        &	1	 &	0	&	0	&	0.5	      &	0.0085(4)	\\
															
	                    &	Rh	             &	1a	        &	1	 &	0	&	0	&	0	      &	0.0052(4)	\\
															
	                    &	Al	             &	4i	        &	1	 &	0	&	0	&	0.1674(2) &	0.0078(5)	\\
															
	                    &	Si	             &	2h	        &	1	 &	0.5	&	0.5	&	0.3569(2) &	0.0075(5)	\\
		\hline													
CeIrAl$_{4}$Si$_{2}$	&	Ce	             &	1b	        &	1	 &	0	&	0	&	0.5	      &	0.0087(3)	\\
															
	                    &	Ir	             &	1a	        &	1	 &	0	&	0	&	0	      &	0.0042(3)	\\
															
	                    &	Al	             &	4i	        &	1	 &	0	&	0	&	0.1662(2) &	0.0066(5)	\\
															
	                    &	Si	             &	2h	        &	1	 &	0.5	&	0.5	&	0.3570(3) &	0.0064(5)	\\
		\hline													
CePtAl$_{4}$Si$_{2}$	&	Ce	             &	1b	        &	1	 &	0	&	0	&	0.5	      &	0.0079(3)	\\
															
	                    &	Pt	             &	1a	        &	1	 &	0	&	0	&	0	      &	0.0052(3)	\\
															
	                    &	Al	             &	4i	        &	1	 &	0	&	0	&	0.1705(2) &	0.0080(6)	\\
															
	                    &	Si	             &	2h	        &	0.868(2)	 &	0.5	&	0.5	&	0.3529(2) &	0.0057(6)	\\

                        &	Pt	             &	2h	        &	0.132(2)	 &	0.5	&	0.5	&	0.3529(2) &	0.0057(6)	\\
\hline
\end{tabular}
  \end{center}
\end{table}

\begin{figure}[H]
\begin{center}
 \includegraphics[scale=.7]{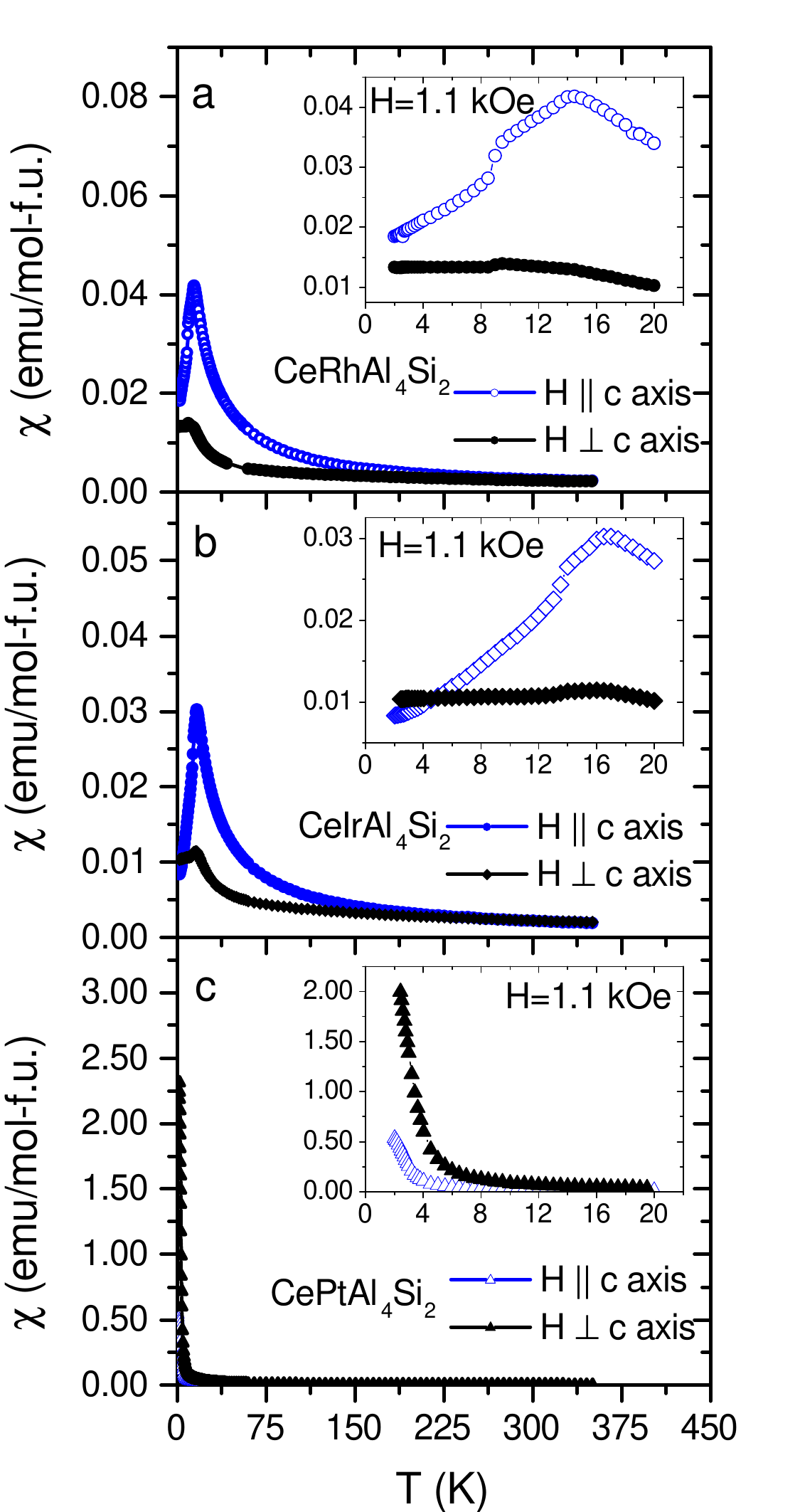}
  \caption{(Color online) Magnetic susceptibility $\chi$ of a) CeRhAl$_{4}$Si$_{2}$, b) CeIrAl$_{4}$Si$_{2}$, and c) CePtAl$_{4}$Si$_{2}$ measured in an applied field of 1.1 kOe. The inset of each panel shows $\chi$ below 20 K.}\label{fig2}
  \end{center}
\end{figure}

The magnetic susceptibility ($\chi \equiv M/H$) measured in a magnetic field of \emph{H} = 1.1 kOe applied parallel ($\chi_{c}$) and perpendicular ($\chi_{ab}$) to the crystallographic \emph{c} axis are plotted in Fig. \ref{fig2}. Anomalies in $\chi$ indicate that CeRhAl$_{4}$Si$_{2}$ and CeIrAl$_{4}$Si$_{2}$ order antiferromagnetically below a Neel temperature ($T_{N}$) of  14 K and 16 K, respectively. Both of these materials show a second transition at a lower temperature - 9 K in CeRhAl$_{4}$Si$_{2}$ and 14 K in CeIrAl$_{4}$Si$_{2}$, suggesting a reorientation of the Ce moments to a new magnetic structure. In CeRhAl$_{4}$Si$_{2}$,  $\chi_c > \chi_{ab}$ down to 1.8 K indicating that \emph{c} axis is the easy magnetic axis. Similar behavior is observed in CeIrAl$_{4}$Si$_{2}$ down to about 5 K, below which the $\chi_{c}$ becomes smaller than  $\chi_{ab}$. The magnetic susceptibility of CePtAl$_{4}$Si$_{2}$ shows a large increase below 5 K reflecting the onset of  ferromagnetism below $T_C=3$ K. In CePtAl$_{4}$Si$_{2}$,  $\chi_{ab}$ is greater than $\chi_{c}$ indicating that the magnetic moments lie in the \emph{ab} plane, opposite to the observed behavior of CeMAl$_{4}$Si$_{2}$ (M=Rh, Ir).

\begin{figure}[H]
\begin{center}
  \includegraphics[scale=.6]{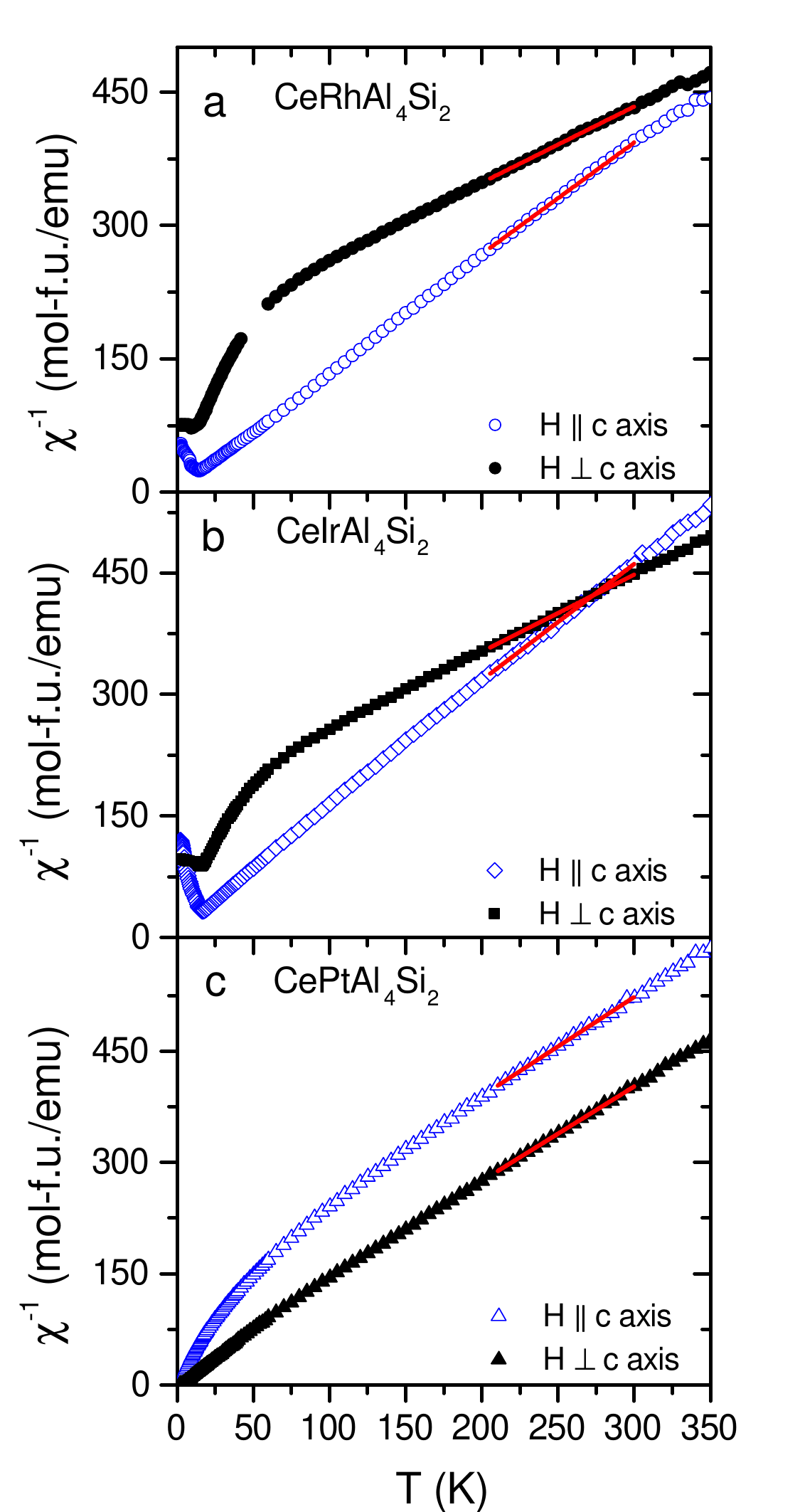}
  \caption{(Color online) Inverse magnetic susceptibility $\chi^{-1}$ of a) CeRhAl$_{4}$Si$_{2}$, b) CeIrAl$_{4}$Si$_{2}$, and c) CePtAl$_{4}$Si$_{2}$. The solid line is a Curie-Weiss fit to the data.}\label{fig3}
  \end{center}
\end{figure}

The inverse magnetic susceptibility ($\chi^{-1}$) of CeMAl$_{4}$Si$_{2}$  (M=Rh, Ir, Pt) is displayed in Fig. \ref{fig3}a-c. In all the three compounds, $\chi^{-1}$  exhibits a linear temperature dependence between 200-300 K. A Curie-Weiss fit to the data of the form $\chi^{-1}$  = C/(\emph{T}-$\theta_{CW}$), where $C=N_A \mu_{eff}^2/3 k_B$ is the Curie constant and $\theta_{CW}$ is the Curie-Weiss temperature, was used to determine the effective moment ($\mu_{eff}$) and  $\theta_{CW}$. The obtained values are presented in Table \ref{table3}. The polycrystalline average of the effective moment, obtained from a Curie-Weiss fit to the polycrystalline average magnetic susceptibility $\chi_{poly} = (2/3)\chi_a + (1/3)\chi_c$,  for CeRhAl$_{4}$Si$_{2}$, CeIrAl$_{4}$Si$_{2}$ and CePtAl$_{4}$Si$_{2}$ is 2.88, 2.72 and 2.49 $\mu{_B}$, respectively, indicating that in all three compounds, Ce is in a Ce$^{3+}$ state for which the expected effective moment is 2.54 $\mu_{B}$. The Curie-Weiss temperatures are negative indicating dominant antiferromangetic interactions, which are characteristic of Ce-based materials with significant Kondo interactions. The anisotropy of  $\theta_{CW}$ may be due to anisotropic magnetic interactions or crystalline electric field effects, which split the Ce$^{3+}$ $J=5/2$ multiplet into 3 doublets. The crystal field effect may also be responsible for the larger effective moments than the Ce$^{3+}$ free ion value in CeRhAl$_{4}$Si$_{2}$ and CeIrAl$_{4}$Si$_{2}$ obtained from $\chi_{ab}$, which have relatively larger and negative values of $\theta_{CW}$ \cite{Maekawa1985}.

\begin{table}[ht]
\caption{Effective magnetic moment ($\mu_{eff}$) and Curie-Weiss temperature ($\theta_{CW}$) of CeMAl$_{4}$Si$_{2}$ compounds obtained from a Curie-Weiss fit to the high temperature inverse susceptibility data as indicated by the red lines in Fig. \ref{fig3}.}\label{table3}
  \begin{center}
  \begin{tabular}{lccc}
\hline								
Compound	                                             &	CeRhAl$_{4}$Si$_{2}$	&	CeIrAl$_{4}$Si$_{2}$	& CePtAl$_{4}$Si$_{2}$   \\
\hline							
$\mu_{eff}$ (H $ \parallel$ \emph{c} axis) ($\mu_{B}$)   &  2.53                    & 2.36                      & 2.45  \\
$\mu_{eff}$ (H $\perp$ \emph{c} axis)	($\mu_{B}$)      & 3.06                     & 2.90                      & 2.52  \\
$\mu_{eff}$ (poly. avg.) ($\mu_{B}$)                     & 2.88         	        & 2.72                      & 2.49  \\
$\theta_{CW}$ H $ \parallel$ \emph{c} axis (K)           & -14  	                & -22 	                    & -92       \\
$\theta_{CW}$ H $\perp$ \emph{c} axis (K)	             & -207 	                & -171  	                & -16 	      \\
\hline													
\end{tabular}
  \end{center}
\end{table}

\begin{figure}[H]
\begin{center}
  \includegraphics[scale=.61]{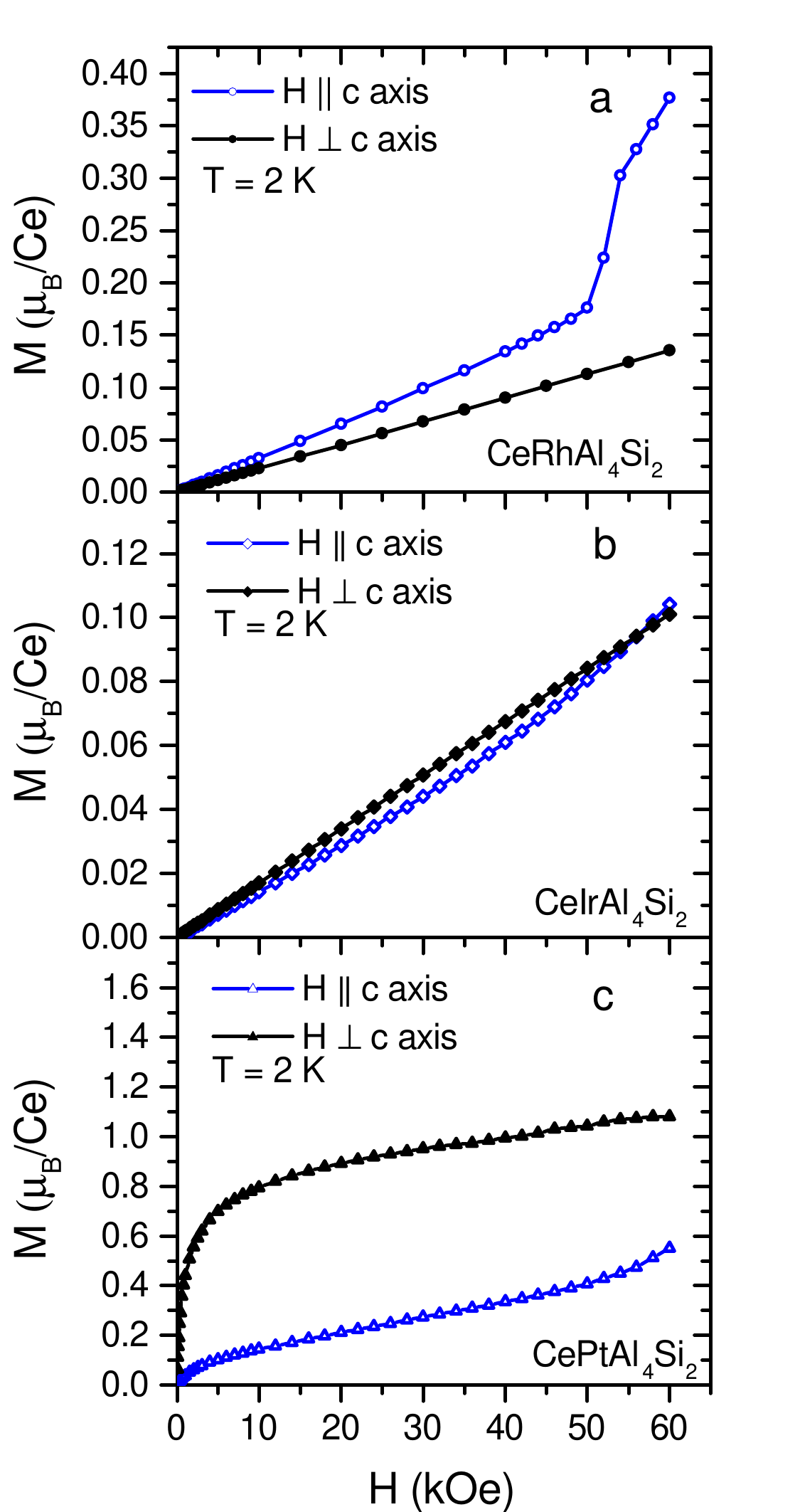}
  \caption{(Color online) \emph{M} versus \emph{H} of a) CeRhAl$_{4}$Si$_{2}$, b) CeIrAl$_{4}$Si$_{2}$, and c) CePtAl$_{4}$Si$_{2}$ measured at T = 2 K.}\label{fig4}
  \end{center}
\end{figure}

Figure \ref{fig4} shows magnetization curves of the three CeMAl$_{4}$Si$_{2}$  compounds at 2 K. In CeRhAl$_{4}$Si$_{2}$  and  CeIrAl$_{4}$Si$_{2}$, \emph{M} vs. \emph{H} shows linear behavior up to 60 kOe when the magnetic field is perpendicular to the \emph{c} axis. However, when the magnetic field is applied parallel to the \emph{c} axis, a metamagnetic transition is observed in CeRhAl$_{4}$Si$_{2}$ at a field of 50 kOe. Although a small curvature is observed in \emph{M} vs. \emph{H} of CeIrAl$_{4}$Si$_{2}$ along the \emph{c} axis, no obvious metamagnetic transition is visible up to 60 kOe. Ferromagnetic behavior is observed in the \emph{M} vs. \emph{H} curves of CePtAl$_{4}$Si$_{2}$ in both directions. The saturation moment for CePtAl$_{4}$Si$_{2}$ for $H\perp c$ is $\mu_{sat}$=0.65 $\mu_{B}$.

\begin{figure}[H]
\begin{center}
  \includegraphics[scale=.73]{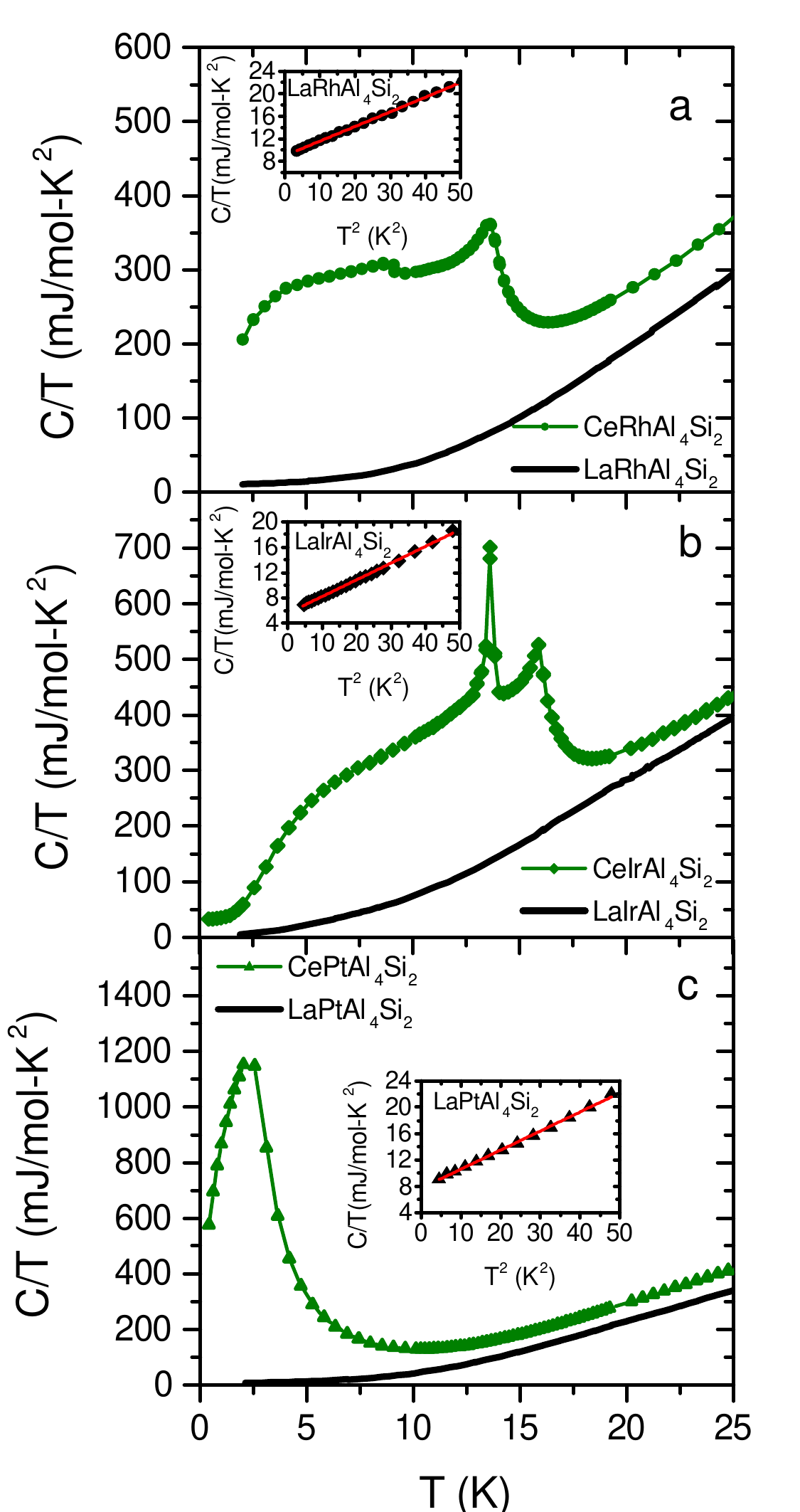}
  \caption{ (Color online) a) Specific heat divided by temperature \emph{C/T} vs \emph{T} of a) CeRhAl$_{4}$Si$_{2}$, b) CeIrAl$_{4}$Si$_{2}$, and c) CePtAl$_{4}$Si$_{2}$. The solid line in each panel shows  \emph{C/T} of the corresponding non-magnetic iso-structural La-analog. The insets show the low temperature fit of the heat capacity of the La-analogs.}\label{fig5}
  \end{center}
\end{figure}
\begin{figure}[H]
\begin{center}
  \includegraphics[scale=.75]{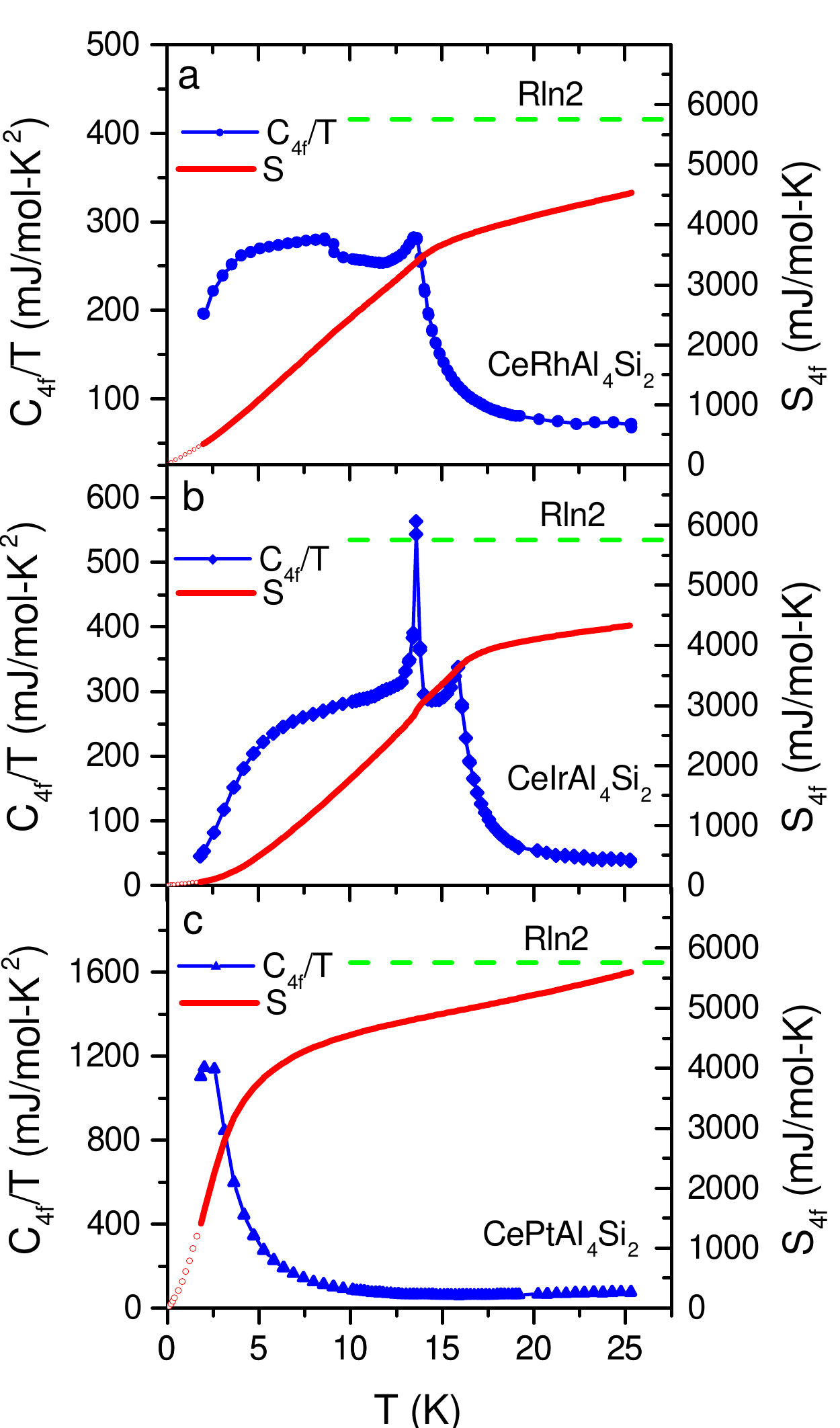}
  \caption{(Color online) $4f$ contribution to the specific heat (left axis) and temperature evolution of the \emph{4f} contribution to the entropy (right axis) of a) CeRhAl$_{4}$Si$_{2}$, b) CeIrAl$_{4}$Si$_{2}$, and c) CePtAl$_{4}$Si$_{2}$. The dotted lines are the extrapolation of \emph{C/T} at the low temperature region.}\label{fig6}
  \end{center}
\end{figure}

The long-range magnetic order in all three compounds is confirmed by specific heat measurements. As shown in the \emph{C/T} vs. \emph{T} plots in Fig. \ref{fig5},   anomalies in $C/T$ coincide with cusps observed in the magnetic susceptibility measurements. The anomaly in the lower transition of CeIrAl$_{4}$Si$_{2}$ [Fig. \ref{fig5}(b)] has the characteristic shape of a first order transition; however, careful analysis of the decay curves of the heat pulse used in the quasi-adiabatic method reveals the lack of a latent heat at the transition, confirming its second-order nature. \emph{C/T} vs. \emph{T} of the non-magnetic iso-structural analogs LaMAl$_{4}$Si$_{2}$ is also plotted in Fig. \ref{fig5}. Subtraction of the non-magnetic La-analogs was used to estimate the $4f$ contribution of the heat capacity to the entropy ($S_{4f} = \int (C(Ce) - C(La)/T)) dT$ as presented in Fig. \ref{fig6}. The electronic contribution to the specific heat $\gamma$, obtained from a linear fit to the \emph{C/T} vs. $T^{2}$ data below 7 K of the form \emph{C/T} = $\gamma$ + $\beta T^2$ (insets in Fig. \ref{fig5}), are 9.0, 5.6 and 7.7 ($mJmol^{-1}K^{-2}$) for LaRhAl$_{4}$Si$_{2}$, LaIrAl$_{4}$Si$_{2}$ and LaPtAl$_{4}$Si$_{2}$, respectively. The respective phonon specific heat coefficient $\beta$ is 0.26, 0.26 and 0.29 ($mJmol^{-1}K^{-4}$). The Debye temperature estimated from $\beta$ using the relation $\theta_{D}$ =$\sqrt[3]{12\pi^{4}rR/5\beta}$, (\emph{r} is number of atoms in the formula unit and \emph{R} is the universal gas constant) for LaRhAl$_{4}$Si$_{2}$, LaIrAl$_{4}$Si$_{2}$ and LaPtAl$_{4}$Si$_{2}$ is 391, 391 and 377 K, respectively. The solid lines in Fig. \ref{fig6} show the temperature dependence of the 4\emph{f} contribution to the entropy $S_{4f}$. The entropy at the magnetic transition temperatures of CeRhAl$_{4}$Si$_{2}$ and CeIrAl$_{4}$Si$_{2}$ and CePtAl$_{4}$Si$_{2}$ is $\sim$0.60, 0.64 and 0.48 of \emph{R}ln(2), respectively. In a simple model, the Kondo temperature $T_{K}$ corresponds to the temperature at which the magnetic entropy reaches 0.5\emph{R}ln(2), which implies that $T_{K}$ is comparable to the ordering temperature in each of these compounds.

\begin{figure}[H]
\begin{center}
  \includegraphics[scale=.7]{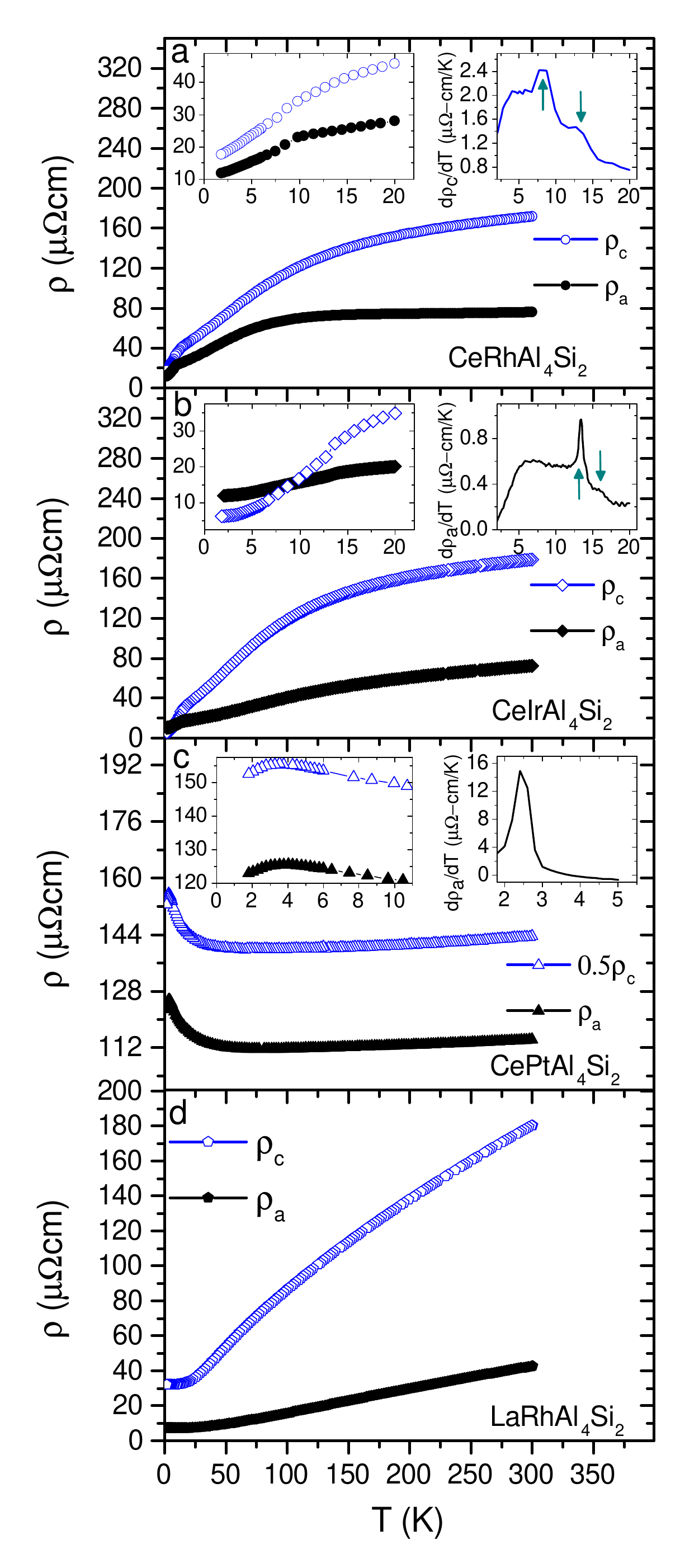}
  \caption{(Color online) Electrical resistivity  as a function of temperature with current applied along the \emph{c} axis ($\rho_{c}$) and along the \emph{a}-axis ($\rho_{a}$) of a) CeRhAl$_{4}$Si$_{2}$, b) CeIrAl$_{4}$Si$_{2}$, c) CePtAl$_{4}$Si$_{2}$, and d) LaRhAl$_{4}$Si$_{2}$. The insets on the left show the data below 12-20 K near the magnetic transitions. The insets on the right show the temperature derivative of the resistivity ($d\rho/dT$) vs \emph{T} below 5-20 K.}\label{fig7}
  \end{center}
\end{figure}

The electrical resistivity for current applied along \emph{c}-axis ($\rho_{c}$) and along the \emph{a}-axis ($\rho_{a}$) is shown in Fig. \ref{fig7}. For CeRhAl$_{4}$Si$_{2}$ and CeIrAl$_{4}$Si$_{2}$,  the resistivity is weakly temperature dependent above 150 K. Below  100 K, the resistivity decreases rapidly with decreasing temperature reflecting either a reduction in scattering from an excited crystal field level or the onset of Kondo coherence of the lattice of Ce ions. In  CePtAl$_{4}$Si$_{2}$, the resistivity is weakly temperature dependent down to 50 K, below which it increases with decreasing temperature and drops after the onset of magnetism. This weak temperature dependence may be a consequence of Si/Pt disorder on the 2h site. For CeRhAl$_{4}$Si$_{2}$ and CeIrAl$_{4}$Si$_{2}$, anomalies in $\rho$ are observed at the two ordering temperatures; a weak anomaly reflects the onset of the upper and a pronounced kink appears at the lower transition in both compounds (inset of Fig. \ref{fig7}a,b).  The residual resistivity ratio (RRR) defined as $\rho_{300K}$/$\rho_{2K}$ in $\rho_{c}$ is about 10 in CeRhAl$_{4}$Si$_{2}$ and 28 in CeIrAl$_{4}$Si$_{2}$.

\begin{figure}[ht]
\begin{center}
  \includegraphics[scale=.4]{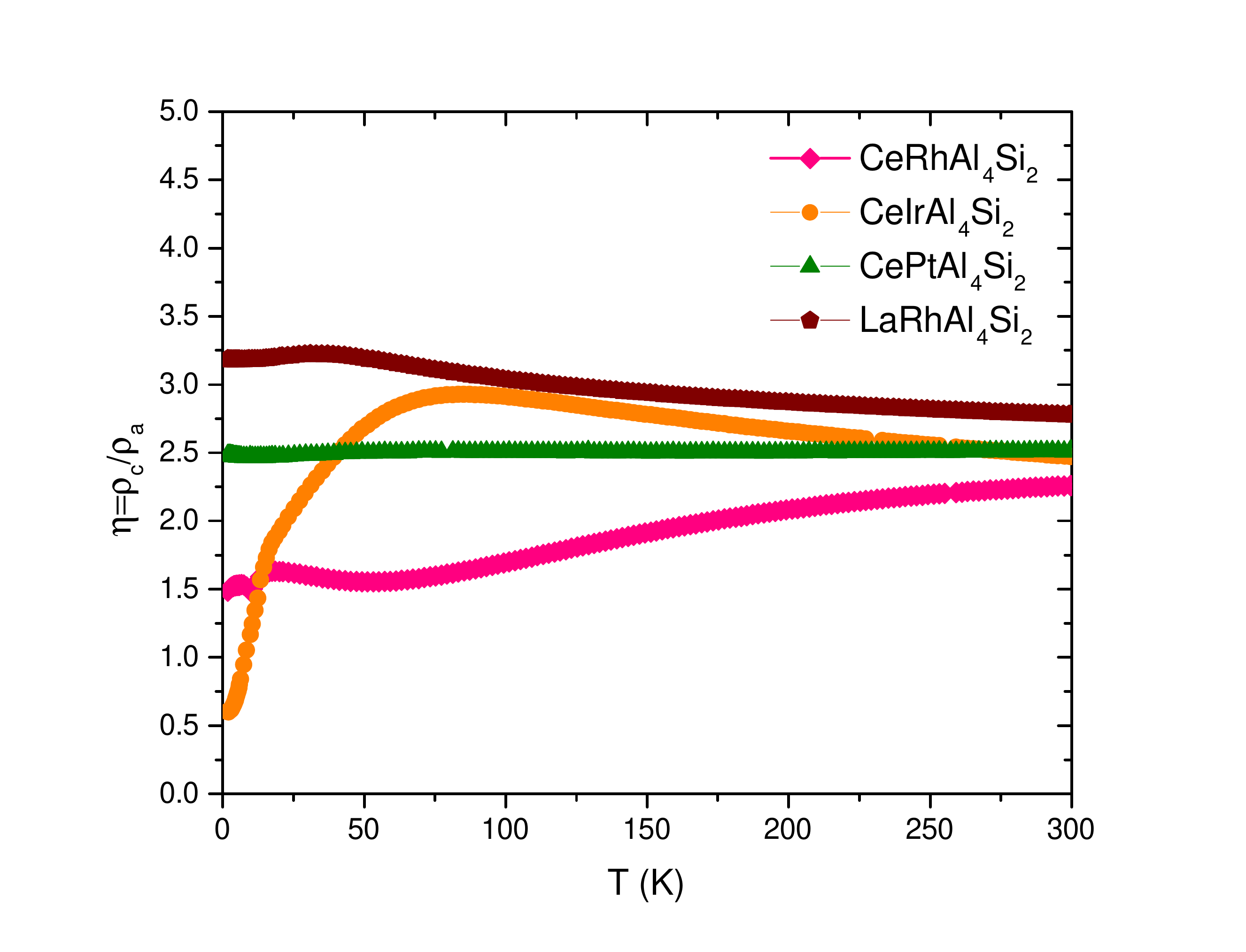}
  \caption{(Color online) The ratio of the \emph{c}-axis to \emph{a}-axis resistivity ($\rho_{c}$/$\rho_{a}$) as a function of temperature.}\label{fig8}
  \end{center}
\end{figure}

Figure \ref{fig8} shows the ratio of the out-of-plane to the in-plane resistivity ($\rho_{c}$/$\rho_{ab}$) as a function of temperature, which is a measure of the electronic anisotropy. The electronic anisotropy $\eta$ is almost same ($\sim$ 2.5) for all the three compounds at room temperature. The large electronic anisotropy of CeMAl$_{4}$Si$_{2}$ and LaRhAl$_{4}$Si$_{2}$ (Fig. \ref{fig8}) suggests that  these materials may have  two-dimensional Fermi surfaces, as discussed in detail below. The anisotropy of both LaRhAl$_{4}$Si$_{2}$ and CePtAl$_{4}$Si$_{2}$ show little change with temperature. On the other hand, the anisotropy   of CeRhAl$_{4}$Si$_{2}$ decreases with decreasing temperature, reaching a value $\eta =1.5$ in the magnetic state.  The anisotropy of  CeIrAl$_{4}$Si$_{2}$ exhibits the largest temperature dependence, with $\eta$ dropping below 1 in the magnetic state. The electronic anisotropy behavior is somewhat similar to the behavior observed in the magnetic susceptibility measurements [Fig.\ref{fig2}], where $\chi_{c}$ always remains greater than $\chi_{ab}$ in CeRhAl$_{4}$Si$_{2}$. In both of these compounds, the antiferromagnetic order influences the anisotropy. In CePtAl$_{4}$Si$_{2}$, there is ferromagnetic ordering with moments lying in the \emph{ab} plane.

\begin{figure}[H]
\begin{center}
  \includegraphics[scale=.77]{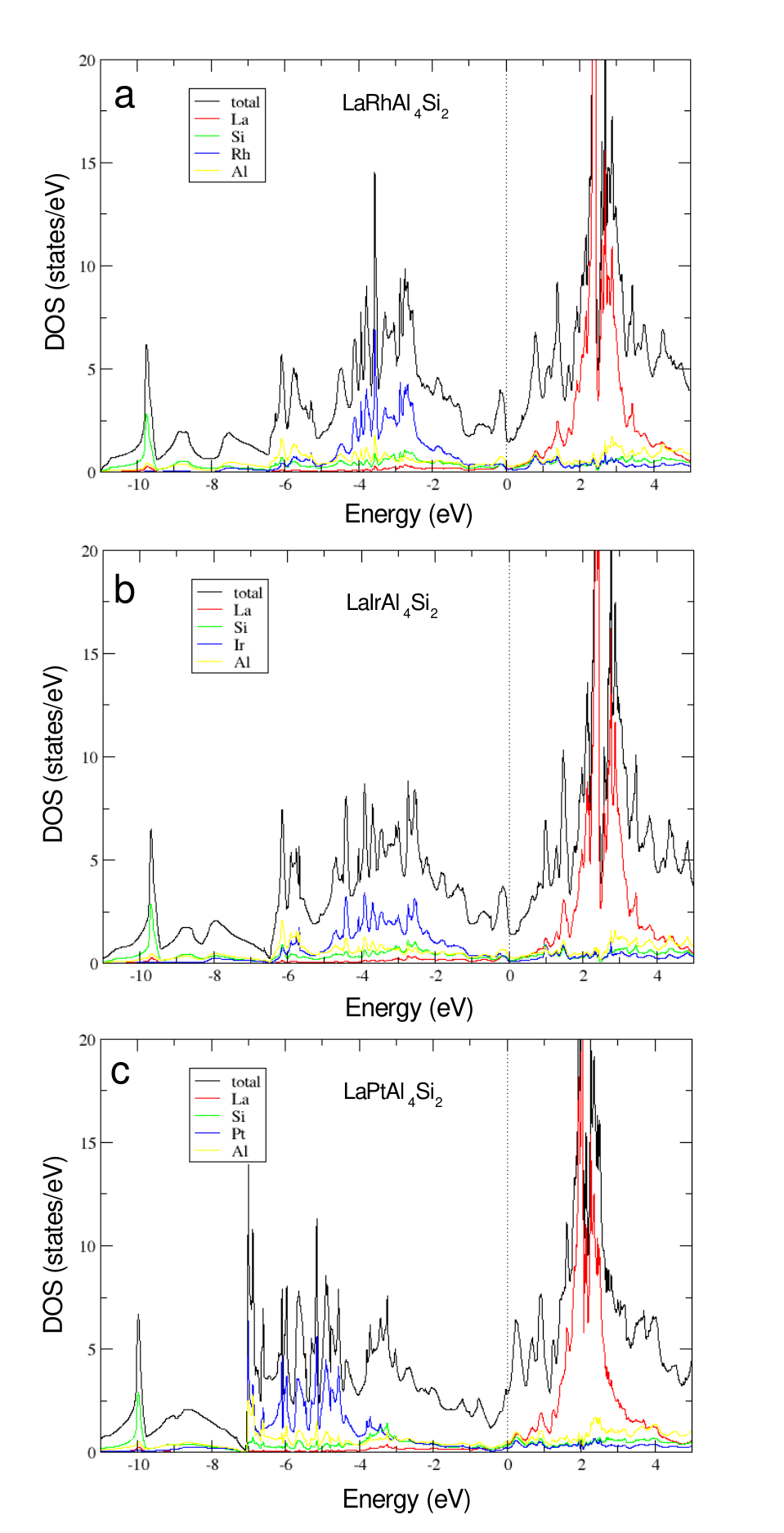}
  \caption{(Color online) Electronic density of states of a) LaRhAl$_{4}$Si$_{2}$, b) LaIrAl$_{4}$Si$_{2}$, and c) LaPtAl$_{4}$Si$_{2}$.}\label{fig9}
  \end{center}
\end{figure}

\begin{figure}[H]
\begin{center}
  \includegraphics[scale=1.2]{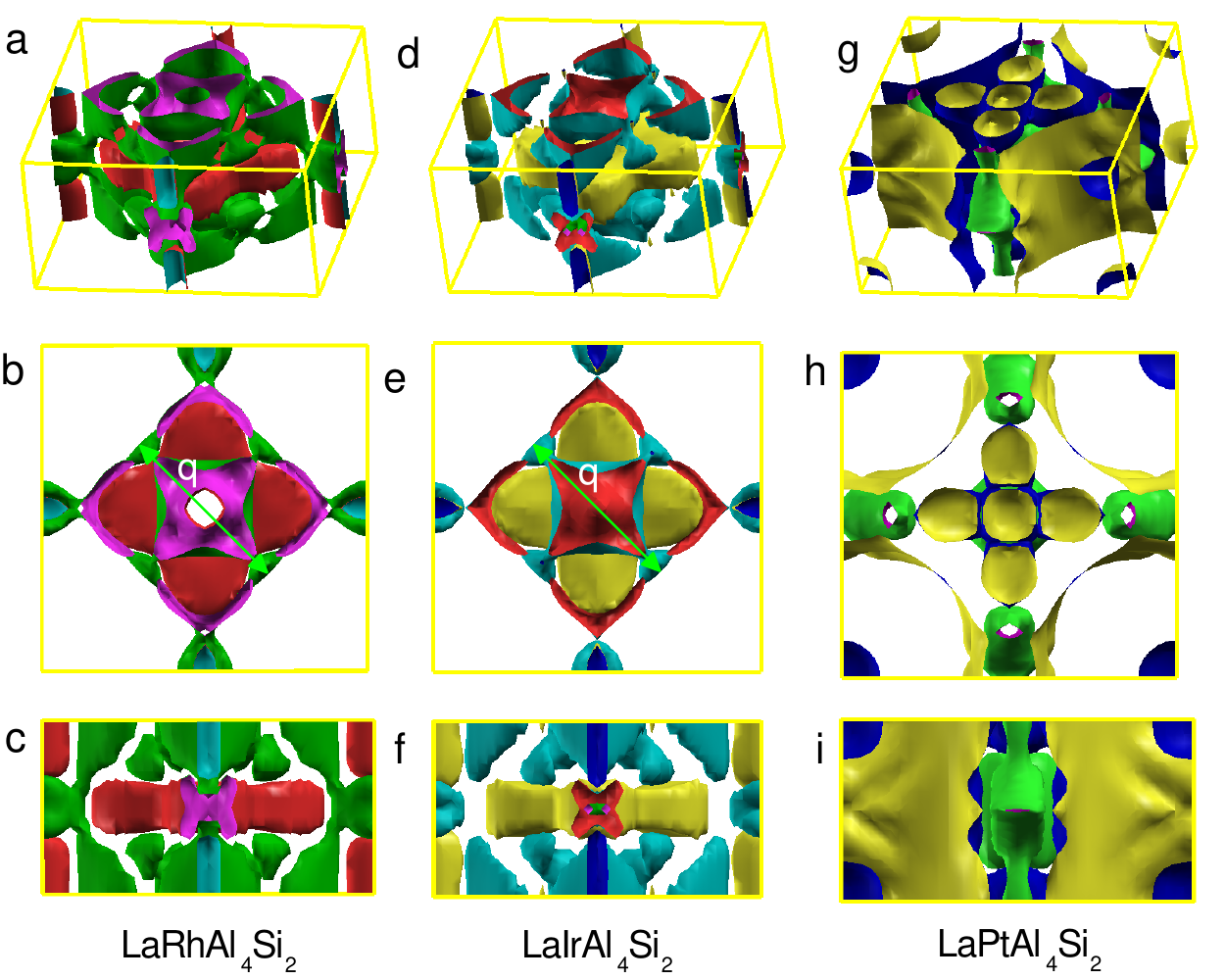}
  \caption{(Color online) Fermi surfaces of (a-c) LaRhAl$_{4}$Si$_{2}$, (d-f) LaIrAl$_{4}$Si$_{2}$, and  (g-i) LaPtAl$_{4}$Si$_{2}$. Projection of the Fermi surface of the respective compounds along \emph{c} axis (b,e,h) and along \emph{a} axis (c,f,i). The arrows in b and e represent the possible nesting vector ($q$).}\label{fig10}
  \end{center}
\end{figure}

To understand the electronic structure  of the RMAl$_{4}$Si$_{2}$ compounds, density functional theory calculations using the generalized gradient approximation were performed. Due to the large magnetic moment in the ferromagnetic state of CePtAl$_{4}$Si$_{2}$ and the small Sommerfeld coefficient of the heat capacity, we assume the Ce $4f$-electrons are localized in these materials. Consequently, the computed Fermi surface of the La analogs provides a reasonable approximation to the actual Fermi surface of the Ce-based compounds at ambient pressure, although the many-body effects responsible for the mass renormalization and magnetic ordering will not be captured \cite{Onuki2012}. The site disorder found in LaPtAl$_{4}$Si$_{2}$ is ignored for the purposes of the calculation. The calculated electronic density of states (DOS) of the three LaMAl$_{4}$Si$_{2}$ is plotted in Fig. \ref{fig9}. The computed density of states at the Fermi level for the three compounds are $N(E_F)$ =3.0, 2.5, and 2.8 states/eV/f.u. for the Rh, Ir and Pt analogs, respectively. This corresponds to a Sommerfeld coefficient  $\gamma$ =7.1, 5.9, and 6.6 mJ/mol K$^2$. The experimental values for LaRhAl$_{4}$Si$_{2}$ and LaPtAl$_{4}$Si$_{2}$ are 1.27 and 1.17 times larger than the computed values suggesting electron-phonon coupling constants of $\lambda$=0.25 and 0.18, respectiviely. However, LaIrAl$_{4}$Si$_{2}$ has a smaller experimental value than the theoretically calculated one. From the density of states plot,  this may result if the measured compound is slightly hole doped with respect to the stoichiometric compound. In Fig. \ref{fig10}, the theoretically calculated Fermi surfaces are displayed for LaMAl$_{4}$Si$_{2}$.  Overall, there is quasi-2D character to the Fermi surface and rather strong nesting present in the Rh and Ir analogs, which can be seen by the z-projection shown in Fig. \ref{fig10}(b,e). Additional nesting also occurs along the c-axis as shown in Fig. \ref{fig10}(c,f). The estimated nesting vector \emph{q} in the  \emph{ab}-plane for LaRhAl$_{4}$Si$_{2}$j and LaIrAl$_{4}$Si$_{2}$ is (0.40$a^{*}$, 0.40$a^{*}$, $q_{z}$) and (0.39$a^{*}$, 0.39$a^{*}$, $q_{z}$), respectively, where $a^{*}$ = $2\pi/a$. We expect that some of the magnetic orderings observed in CeRhAl$_{4}$Si$_{2}$ and CeIrAl$_{4}$Si$_{2}$ are due to these nesting wave vectors. Finally, LaPtAl$_{4}$Si$_{2}$ has one additional electron relative to the Rh and Ir compounds, which leads to a qualitatively different Fermi surface, although the strong quasi-2D character remains. This Fermi surface of LaPtAl$_{4}$Si$_{2}$ is expected to be similar to that of CeRhAl$_{4}$Si$_{2}$ and CeIrAl$_{4}$Si$_{2}$ under high pressure, in which the Ce-4f electron is delocalized and is incorporated into the Fermi volume.  DFT calculations on CeRhAl$_{4}$Si$_{2}$ (not shown) where the Ce-4f electron is treated as delocalized give very similar Fermi surfaces as found in LaPtAl$_{4}$Si$_{2}$. It is intriguing that the  Fermi surface of LaPtAl$_{4}$Si$_{2}$ is somewhat similar to the well known Fermi surface found in overdoped cuprates \cite{Damascelli2003}.  It will be very interesting to see if superconductivity occurs in the CeMAl$_{4}$Si$_{2}$ compounds under applied pressure.

%as evidenced by the small Sommerfeld coefficient of the heat capacity extrapolated from above the ordering temperature, and because a large fraction of the entropy is recovered by the magnetic ordering temperature (Fig. 6).

The CeMAl$_{4}$Si$_{2}$ materials present an opportunity for exploration of strongly correlated electron behavior beyond the magnetic ground states described herein.  The competition of Kondo and RKKY interactions was described by Doniach \cite{Doniach1977}.  For small values of the exchange interaction between the f-electron/conduction-electron $\mathcal{J} \propto - V_{kf}^2/|\epsilon_{f} - E_F|$, where  $V_{kf}$  is the hybridization, $\epsilon_{f}$ is the energy of the $f$-level and  $E_F$ is the Fermi level, the RKKY interaction ($T_{RKKY} \propto \mathcal{J}^2 N(E_F)$) dominates, resulting in  a magnetically ordered ground state.  As the Kondo interaction ($T_{K} \propto exp[-1/\mathcal{J}N(E_F)$]) increases, the ordering temperature passes through a maximum, then is suppressed to zero temperature at the quantum critical point.  Based upon the high ordering temperature of CeRhAl$_{4}$Si$_{2}$ and CeIrAl$_{4}$Si$_{2}$, these compounds might be placed near the maximum on the dome of magnetism in the Doniach diagram, consistent with the estimate that $T_{K}$ is comparable to $T_{N}$.  The application of pressure moves the system from left to right on the Doniach diagram since the smaller Ce$^{4+}$ ($4f^{0}$) state is preferred over the larger Ce$^{3+}$ ($4f^{1}$) state, leading to a decrease of $T_N$ with $P$. Thus, measurements on CeMAl$_{4}$Si$_{2}$ (M=Rh, Ir) under pressure are a particularly promising avenue to search for interesting correlated electron behavior.   One might be tempted to place CePtAl$_{4}$Si$_{2}$ on the far left of the Donaich diagram, given the large ordered moment in the ferromagnetic state and the small (extrapolated) $\gamma$ value ($\sim$ 50 mJ/mol K$^2$), but the small amount of entropy at $T_C$  (0.48 Rln(2)) suggests that the Kondo interaction may be significant in this material as well.  Quantum oscillation measurements to probe the 2D nature of the Fermi surface and the $4f$ contribution to the Fermi volume in the ground state at the lowest temperature would also be helpful to understand the Ce $4f$ electrons in CeMAl$_{4}$Si$_{2}$.

\section[S4]{Conclusion}

In summary, the physical properties including the crystal structure, specific heat, electrical resistivity, magnetiztion, and magnetic susceptility of the tetragonal RMAl$_{4}$Si$_{2}$ (R=La, Ce; M=Rh, Ir, Pt) materials are reported.   Both CeRhAl$_{4}$Si$_{2}$ and CeIrAl$_{4}$Si$_{2}$ exhibit two antiferromagnetic transitions at relatively high temperature ($\sim$ 10-15 K), while CePtAl$_{4}$Si$_{2}$ orders ferromagnetically at $T_C=3$ K. Electronic structure calculations reveal quasi-2D character of the Fermi surface for all three LaMAl$_{4}$Si$_{2}$ analogs with a propensity for nesting, although the Fermi surface is qualitatively different for LaPtAl$_{4}$Si$_{2}$ compared to  LaRhAl$_{4}$Si$_{2}$ and LaIrAl$_{4}$Si$_{2}$. Measurements under pressure are planned to search for quantum criticality in these CeMAl$_{4}$Si$_{2}$  materials.
\\
\\
{\bf{Acknowledgements}}
\\
\\
Work at Los Alamos National Laboratory was performed under the auspices of the US Department of Energy, Office of Basic Energy Sciences, Division of Materials Sciences and Engineering, and PECASE funding from the US DOE, OBES, Division of Material Science and Engineering. The EDS measurements were performed at the Center for Integrated Nanotechnologies, an Office of Science User Facility operated for the U.S. Department of Energy (DOE) Office of Science. Los Alamos National Laboratory, an affirmative action equal opportunity employer, is operated by Los Alamos National Security, LLC, for the National Nuclear Security Administration of the U.S. Department of Energy under contract DE-AC52-06NA25396.

\section*{References}
\bibliographystyle{ieeetr}

%\bibliography{library}

\end{document}